\documentclass[sigconf]{acmart}
\usepackage{hyperref}
\usepackage{booktabs} 
\usepackage{multirow}
\usepackage[normalem]{ulem}
\useunder{\uline}{\ul}{}
\usepackage{graphicx}
\usepackage{algorithm}
\usepackage{algorithmicx}
\usepackage{algpseudocode}

\usepackage{subfigure}
\usepackage{enumitem}
\usepackage{bbding}
\AtBeginDocument{%
  \providecommand\BibTeX{{%
    \normalfont B\kern-0.5em{\scshape i\kern-0.25em b}\kern-0.8em\TeX}}}
\setcopyright{acmcopyright}
\copyrightyear{2019} 
\acmYear{2019} 
\acmConference[KDD'20]{ACM Conference}{Aug 22--27 2020}{San Diego, California}
\acmBooktitle{ACM Conference (KDD'20), Aug 22--27, 2020, San Diego, California}
\acmPrice{15.00}
\acmDOI{10.1145/3298689.3347033}
\acmISBN{978-1-4503-6243-6/19/09}

\begin{document}

\title{A Practical Incremental Method to Train Deep CTR Models}
\author{Yichao Wang, Huifeng Guo, Ruiming Tang, Zhirong Liu, Xiuqiang He} 
\affiliation{Noah's Ark Lab, Huawei, China.}
\email{{wangyichao5, huifeng.guo, tangruiming, liuzhirong, hexiuqiang1}@huawei.com}



\begin{abstract}
Deep learning models in recommender systems are usually trained in the batch mode, namely iteratively trained on a fixed-size window of training data. 
Such batch mode training of deep learning models suffers from low training efficiency, which may lead to performance degradation when the model is not produced on time.
To tackle this issue, incremental learning is proposed and has received much attention recently. Incremental learning has great potential in recommender systems, as two consecutive window of training data overlap most of the volume. It aims to update the model incrementally with only the newly incoming samples from the timestamp when the model is updated last time, which is much more efficient than the batch mode training. 
However, most of the incremental learning methods focus on the research area of image recognition where new tasks or classes are learned over time. In this work, we introduce a practical incremental method to train deep CTR models, which consists of three decoupled modules (namely, data, feature and model module). Our method can achieve comparable performance to the conventional batch mode training with much better training efficiency. We conduct extensive experiments on a public benchmark and a private dataset to demonstrate the effectiveness of our proposed method.
\end{abstract}


\maketitle
\section{Introduction}

Internet users can access a huge number of online products and services, therefore it becomes difficult for users to identify what might interest them. To reduce information overload and to satisfy the diverse needs of users, personalized recommender systems are playing an important role in modern society. Accurate personalized recommender systems benefit both demand-side and supply-side including publisher and platform. 

Click-Through Rate (CTR) prediction is to estimate the probability that a user will click on a recommended item under a specific context. It plays a crucial role in personalized recommender system, especially in the business of app store and online advertising. Nowadays, deep learning approaches have attracted more and more attention due to the superiority over prediction performance and automated feature exploration. Therefore, many industrial companies deploy deep CTR models in their recommender system, such as Wide \& Deep~\cite{cheng2016wide} in Google Play, DeepFM~\cite{deepfm} and PIN~\cite{pin} in Huawei AppGallery, DIN~\cite{din} and DIEN~\cite{dien} in Taobao, etc.

\begin{figure}[h]
  \centering
  \includegraphics[width=0.8\linewidth]{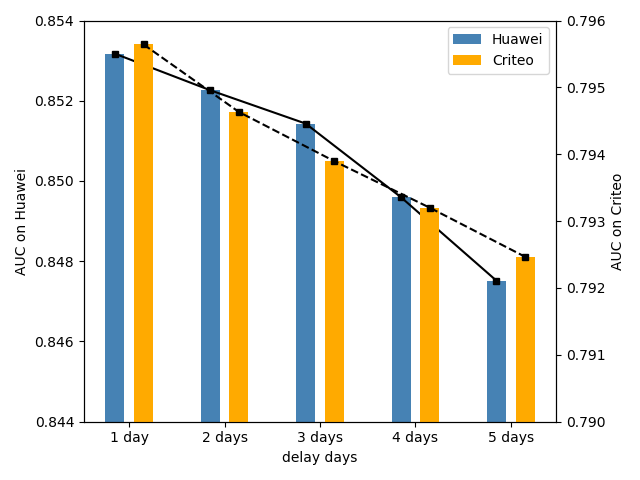}
  \caption{Model performance degrades when the model stops updating for different days. X-axis is different gaps between training set and test set.}
  \label{fig:mean_auc_huawei}
\end{figure}

However, every coin has two sides. To achieve good performance, Deep CTR models with complicated architectures need to be trained on huge volume of training data for several epochs, therefore they all suffer from low training efficiency. Such low training efficiency (namely, long training time) may lead to performance degrade when the model is not produced on time. We observe such performance degradation when the model stops updating in app recommendation scenarios in Huawei AppGallery, as presented in Figure~\ref{fig:mean_auc_huawei}. For instance, if the model stops updating for 5 days, the model performance degrades 0.66\% in terms of AUC, which would lead to significant loss of revenue and also user experience. 
Hence, as can be observed, how to improve training efficiency of Deep CTR models without hurting its performance is an essential problem in recommender system. 
Distributed learning and incremental learning are two common paradigms to tackle this problem from different perspectives. Distributed learning requires extra computational resources and distributes training data and the model to multiple nodes to accelerate training. On the other side, incremental learning changes the training procedure from batch mode to increment mode, which utilizes just the most recent data to update the current model. However, most deep models in industrial recommender system are trained in the batch mode, where a fixed-size window of training data (usually in a multi-billion scale) is used to train the model iteratively.  In this work, we focus on devising incremental method to train deep CTR models, which aims to improve the training efficiency significantly without degrading the model performance.

\begin{figure*}
    \centering
    \includegraphics[width=\linewidth]{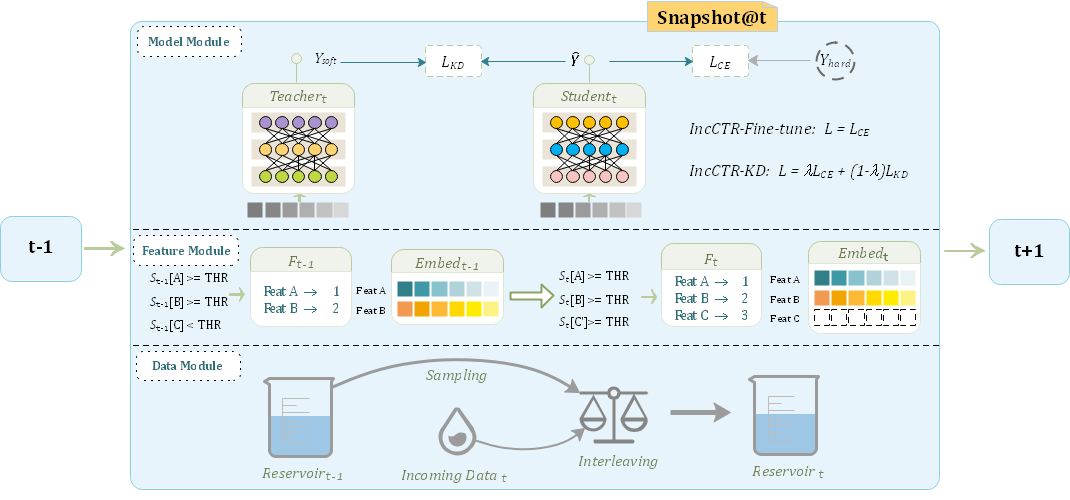}
    \caption{Overview of IncCTR architecture, where t indicates the incremental step.}
    \label{fig:architecture}
\end{figure*}


However, to the best of our knowledge, most of the existing incremental learning methods mainly concentrate on image recognition field where new tasks or classes are learned over time. While incremental learning for deep CTR models faces different circumstances from image recognition, such as incoming new features, etc, therefore there is a need to look into this topic seriously. In this paper, we propose a practical incremental method \textit{IncCTR} for deep CTR models. As presented in Figure~\ref{fig:architecture}, three decoupled modules are integrated in our model: \textit{Data Module}, \textit{Feature Module} and \textit{Model Module}. Data module mimics the functionality of a reservoir, constructing training data from both historical data and incoming data. Feature module is designed to handle new features from incoming data and initialize both existing features and new features wisely. Model module employs knowledge distillation to fine-tune the model parameters, balancing learning knowledge from the previous model and from incoming data. More specifically, we look into two different choices for the teacher model. 

The main contributions of this work are listed as follows:
\begin{itemize}
    \item We highlight the necessity of incremental learning in recommender system through rigorous offline simulations. We propose a practical incremental method, IncCTR, to train deep CTR models.
    \item IncCTR consists of data module, feature module and model module, which have the functionality of constructing training data , handling new features and fine-tuning model parameters, respectively. 
    \item We conduct extensive experiments on a public benchmark and a private industrial dataset from Huawei AppGallery to demonstrate that IncCTR is able to achieve comparable performance to the batch mode of training, with significant improvement on training efficiency.  Moreover, ablation study of each module in IncCTR are performed. 
\end{itemize}
The rest of the paper is organized as follows. In Section \ref{sec:preliminary}, we introduce some preliminaries for better understanding our method and application. We elaborate our incremental learning framework IncCTR and three individual modules in detail in Section \ref{sec:methodologies}. In section \ref{sec:experiment}, results of comparison experiments and ablation studies are reported to verify the effectiveness of our proposed framework. Lastly, we draw a conclusion and discuss future work in section \ref{sec:conclusion}.

\section{PRELIMINARIES}
\label{sec:preliminary}
In this section, we introduce some notations, basic knowledge about deep CTR models. Also, the training modes (batch mode and incremental mode) are presented and compared.
\subsection{Deep CTR Model}   
Recently, various deep CTR models are proposed, such as DeepFM~\cite{deepfm}, Wide \& Deep~\cite{cheng2016wide}, PIN~\cite{pin}, DIN~\cite{din}, and DIEN~\cite{dien}. Generally, deep CTR models include three parts: embedding layer, interaction layer, and prediction layer.
\subsubsection{Embedding layer}
In most CTR prediction tasks, data is collected in a multi-field categorical form~\cite{crossnet,Zhang2016Deep,pin}. Each data instance is transformed into a high-dimensional sparse (binary) vector via one-hot encoding~\cite{He2014Practical}. For example,
the raw data instance (Gender=Male, Height=185, Age=18, Name=Bob) can be represented as:
\[
  \underbrace{(0,1)}_{\text{Gender=Male}}\underbrace{ (0,...,1,0,0)}_{\text{Height=185}}\underbrace{(0,1,...,0,0)}_{\text{Age=18}}\underbrace{(1,0,0,..,0)}_{\text{Name=Bob}.}
\]

An embedding layer is applied to compress the raw features to low-dimensional vectors before feeding them into neural networks. For a univalent field, (e.g., ``Gender=Male''), its embedding is the feature embedding; For a multivalent field (e.g., ``Interest=Football, Basketball''), the field embedding is the average of the feature embeddings~\cite{dnnyoutube}. More formally, in an instance, each field $f_i\ (1 \le i \le m)$ is represented as a low-dimensional vector $e_i \in R^{1\times k}$, where $m$ is the number of fields and $k$ is the embedding size.
Therefore, each instance can be represented as an embedding matrix $E = (e_1^\top, e_2^\top, ..., e_{m}^\top)^\top \in R^{m\times k}$. 
Assume there are $n$ features, the embeddings of all the features form an embedding table $\mathcal{E}\in R^{n\times k}$.
\subsubsection{Interaction and Prediction Layers}
The key challenge in CTR prediction is modelling feature interactions. Existing deep CTR models utilize product operation and multi-layer perception (MLP) to model the explicit and implicit feature interactions, respectively. For example, DeepFM~\cite{deepfm} adopts Factorization Machine~\cite{fm} to model the order-2 feature interactions and MLP to model the high-order feature interactions. The method about how to model feature interactions is beyond the scope of this work, those who are interested in such techniques, please refer to~\cite{deepfm,xdeepfm,fgcnn,pin,DCN}.

After the interaction layer, the prediction $\hat{y}$ is generated as the probability of the user will click on a specific item within such context.
Then, the cross-entropy loss is used as the objective function:
\begin{equation}
    \label{loss_ce}
    \mathcal{L}_{CE}(y,\hat{y}) = -y \log \hat{y} -(1-y) \log (1-\hat{y}).
\end{equation} 
with  $y$ as the label.
\subsection{Batch Mode V.S. Incremental Mode}
In this section, we present and compare two different training modes, namely batch mode and incremental mode. 
\subsubsection{Training with Batch Mode}
The model trained in the batch mode is iteratively learned based on the data from a fixed-size time window. When new data is arriving, the time window slides forward. As shown in Figure~\ref{fig:batch_vs_incre},  ``model 0'' is trained based on data from day 1 to day 10. Then when new data (namely ``day 11'') is arriving, a new model (namely ``model 1'') is trained based on data from day 2 to day 11. For similar procedure, ``model 2'' is trained on data from day 3 to day 12.
\subsubsection{Training with Incremental Mode}
\begin{figure}[h]
  \centering
  \includegraphics[width=0.8\linewidth]{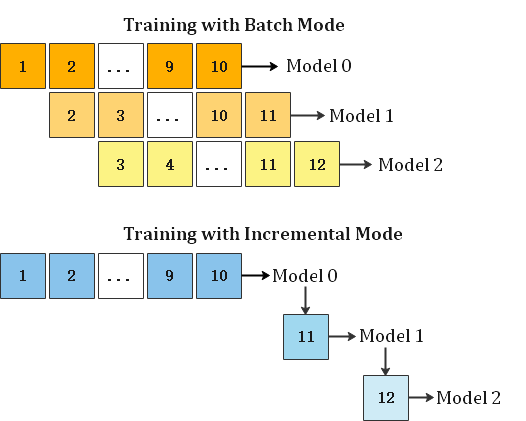}
  \caption{Training with Batch Mode V.S. Incremental Mode. Each block represents one day of training data.}
  \label{fig:batch_vs_incre}
\end{figure}

With incremental mode, the model is trained based on the existing model and new data. As shown in Figure~\ref{fig:batch_vs_incre}, the ``Model 1'' is trained based on the existing model ``Model 0'' (which is trained on data from day 1 to day 10), and data from day 11. Then ``Model 1'' turns into the existing model. Consequently, when data from day 12 arrives, ``Model 2'' is trained based on the ``Model 1'' and data from day 12.

As can be seen, when training with batch mode, two consecutive time window of training data overlap most of the volume. For instance, data from day 1 to day 10 and data from day 2 to day 11 overlap the part from day 2 to day 10, where 80\% of the volume are shared. Under such circumstance, replacing batch mode with incremental mode improves the efficiency significantly, while such replacement is highly possible to retain the performance.  
\section{Methodologies}
\label{sec:methodologies}
An overview of our incremental learning framework IncCTR is shown in Figure~\ref{fig:architecture}. 
Three modules are designed from the perspective of \textit{feature}, \textit{model} and \textit{data} respectively to balance the trade-off between learning from historical data and incoming data. Specifically, data module serves as a reservoir, constructing training data from both historical data and incoming data. Feature module handles new features from incoming data and initializes both existing features and new features properly. Model module employs knowledge distillation to fine-tune the model parameters. 

\subsection{Feature Module}
In recommendation and information retrieval scenarios, the feature dimension is usually very high, i.e., in million- or even billion-scale~\cite{DHPS_baidu}. The occurrence frequency of such large number of features follows long-tailed distribution, where only minor proportion of the features occur frequently and the rest are rarely presented. As observed in~\cite{ftrl}, half of the features in their model occur only once in the whole training data. The features that are rarely occurred are difficult to be learned well. Therefore, when training with batch mode, features are needed to be categorized to ``frequent'' or ``infrequent'' by counting the number of occurrences of each feature. More formally, a feature $x$ with its occurrence $S[x]$ larger than a pre-defined threshold $THR$ (i.e., $S[x] > THR$) is considered as ``frequent'' and is learned as an individual feature. The rest ``infrequent'' features are treated as a special dummy feature \textit{Others}. After such processing, each feature is mapped to a unique id by some policy like auto-increment, hash-coding, and etc. We take an auto-increment policy $F$ for simplicity. 
In batch mode, policy $F$ is constructed from scratch by assigning unique ids to individual features from the training data in a fixed-size window, where the unique ids increase one by one automatically. 

\begin{figure}[h]
    \centering
    \includegraphics[width=0.8\linewidth]{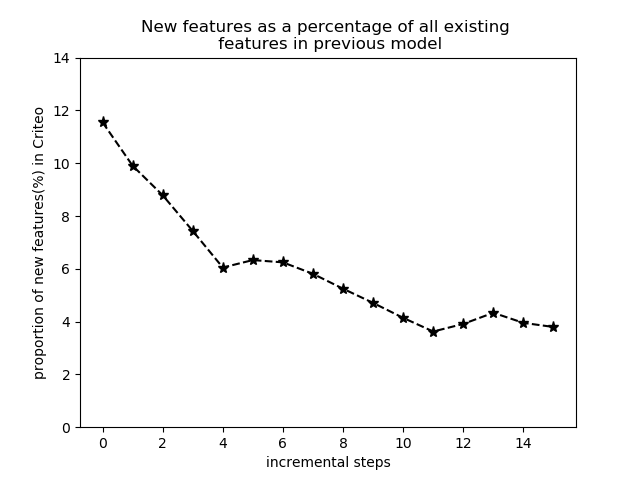}
    \caption{Proportion of new features compared to the set of existing features as blocks of new data comes in, observed from Criteo dataset.
    }
    \label{fig:new_feature_proportion}
\end{figure}

However, training with incremental mode brings additional issue as new features appear when new data comes in. As displayed in Figure~\ref{fig:new_feature_proportion}, every block of new data brings a certain proportion of new features. For instance, as observed from Criteo dataset, the first block of new data imports 12\% of new features compared to the set of existing features before this block, while even the 14th block still brings 4\% of new features. Therefore, the policy $F$ needs to be updated incrementally when new data comes in.
It is possible that a feature $x$, which is previously treated as \textit{Others}, is considered as a unique feature if its occurrence $S[x]$ is above the threshold $THR$ after new data coming in. 

After assigning proper ids to all the features, feature module in IncCTR initializes both existing features and new features. When we start training with batch mode, all the feature embeddings $\mathcal{E}$ are initialized randomly. Whereas, in case of incremental mode, we initialize the embeddings of existing features $\mathcal{E}_{exist}$ and the embeddings of new features $\mathcal{E}_{new}$ separately. 

The functionality of feature module (namely, new feature assignment and feature embedding initialization) is presented in Algorithm~\ref{algo_feature}. When new data comes in, we firstly update the occurrences of each feature (line 3) and inherit the existing policy of feature assignment (line 4). If a feature from the new data is new with its occurrence larger than the threshold (line 6), it is added to the policy with id incremented by one (line 7). Feature embeddings are initialized separately, depending on whether a feature is new or not. For an existing feature, it inherits from the embedding of the existing model as its initialization (line 11). Such inheriting transfers the knowledge in historical data to the model that will be trained incrementally. For a new feature, its embedding is randomly initialized as no prior knowledge is available (line 12).

\subsection{Model Module}
\label{sec:fine-tune-kd}
In this section, we introduce the model module in IncCTR, which trains the model properly, such that the model still ``remembers'' the knowledge from historical data and also makes some ``progress'' from the new data.

\textbf{Fine-tune}.
Besides the embedding of existing features, the network parameters are also inherited from the existing model as warm-start. To fine-tune all the model parameters with incremental mode, we apply some auxiliary tricks to achieve good performance. For instance, we conduct a lower learning rate for $\mathcal{E}_{exist}$ compared to that for $\mathcal{E}_{new}$. The training details of fine-tune are presented in line 19 to line 25 in Algorithm~\ref{algo_incCTR}. The model is optimized by minimizing cross entropy between prediction and groundtruth. We train the model for a fixed number of epochs, where empirically we set the number to be 1 (line 25).
\begin{algorithm}[H]
    \caption{Feature Module: New Feature Assignment and Feature Embedding Initialization}
    \label{algo_feature}
    \begin{algorithmic}[1]
    \Require features of coming data $X_t$; labels of coming data $Y_t$; existing model $M_{t-1}$; existing policy of feature assignment $F_{t-1}$; frequency of existing features $S_{t-1}$
    \Ensure initialized feature embeddings $\mathcal{E}_{t}$
    \State \textbf{Initialize}: feature frequency threshold $THR$
    \State \textbf{New Feature Assignment}:
    \State $S_{t} \leftarrow Feature\_Frequency\_Update(S_{t-1}, X_t)$
    \State $F_{t} \leftarrow F_{t-1}$
    \For{all features $f$ in $X_{t}$}
    \If{$f\notin F_t \wedge S_{t}[f] > THR$}
    \State $F_{t} \leftarrow add(f, |F_t|+1)$
    \EndIf
    \EndFor
    \State \textbf{Feature Embedding Initialization}:
    \State $\mathcal{E}_{exist} \leftarrow \mathcal{E}_{t-1}$  
    \State $\mathcal{E}_{new} \leftarrow Random\_Initialize$
    \State $\mathcal{E}_{t} \leftarrow \mathcal{E}_{exist} \cup \mathcal{E}_{new}$
    \State \textbf{Return}: $\mathcal{E}_{t}$
    \end{algorithmic}
\end{algorithm}
\textbf{Knowledge distillation}.
Beyond ``fine-tune'' presented above, we introduce knowledge distillation (KD) method to enhance the knowledge learned from the historical data (namely, to avoid catastrophic forgetting). 
Hinton et al. in \cite{KD} use KD to transfer knowledge from an ensemble of models into a single model for efficient deployment, where KD loss is used to preserve knowledge from the cumbersome model through encouraging the outputs of distilled model to approximate that of cumbersome model. Similarly, the authors of LwF~\cite{LWF} perform KD to learn new task while keeping knowledge on old tasks. Borrowing the similar idea, KD can also be used in incremental learning scenario to learn new knowledge from incoming data while preserving memory on historical data. 

Several options are available to design the teacher model when we apply KD in IncCTR. We present two such options.
\begin{itemize}
    \item \textbf{KD-batch.} Outdated model trained with batch mode can be a natural choice for teacher model to distill the incremental model, which can preserve performance on the historical data within a fixed-size window. We refer the KD method with such teacher trained with batch mode as ``KD-batch''.
    \item \textbf{KD-self.} As training a model with batch mode as teacher model needs extra computation resources, it is more convenient to perform the previous incremental model as the teacher. In such case, the successive incremental model is trained with the supervision of the previous incremental model. We refer such a design as ``KD-Self''. Similar idea is employed in \emph{BANs} \cite{BANs}, where a consecutive student model is initialized randomly while taught by the previous teacher model, and the ensemble of multiple student generations was conducted to achieve desirable performance. This work is in image recognition field where all the models are trained in batch mode, which is significantly different from our framework.
\end{itemize}

When performing KD, we utilize the soft targets $Y_{soft}$ generated by the teacher model on the incoming data. The objective function is formulated as follows:
\begin{equation}
    \label{loss_final}
    \mathcal{L} = \mathcal{L}_{CE}(Y, \hat{Y}) + \mathcal{L}_{KD}(\hat{Y}, Y_{soft}) + \mathcal{R}
\end{equation}
\begin{equation}
    \label{loss_kd}
    \mathcal{L}_{KD}(Y, Y_{soft}) = \mathcal{L}_{CE}(\sigma(\frac{Z}{\tau}), \sigma(\frac{Z_{soft}}{\tau}))
\end{equation}
\begin{equation}
    \label{loss_ce_sum}
    \mathcal{L}_{CE}(Y, \hat{Y}) = \sum\limits_{y_i\in Y}\mathcal{L}_{CE}(y_i, \hat{y}_i)
\end{equation}
The new objective function combines the standard binary cross-entropy $\mathcal{L}_{CE}(\cdot)$ (where $Y$ and $\hat{Y}$ denote the groundtruth and outputs of the new model respectively) and KD loss $\mathcal{L}_{KD}(\cdot)$. KD loss $\mathcal{L}_{KD}(\cdot)$ is the cross entropy between $\hat{Y}$ and $Y_{soft}$ (where $Y_{soft}$ is the prediction of the teacher model) which are computed based on logits $Z$ and $Z_{soft}$ of two models. Temperature $\tau$ is applied to get soft targets and $\mathcal{R}$ is the regularization term. The intuition of loss function in Equation~\ref{loss_final} is that the knowledge of the distilled model should be precise to the new data (first term), while it should not be significantly different from the knowledge of the teacher model (second term).

The training details of KD-batch and KD-self are presented in line 3 to line 5 and line 11 to line 17 in Algorithm~\ref{algo_incCTR}. The difference between KD-batch and KD-self is how the teacher model $Teacher_t$ is trained. Remind that the teacher model in KD-batch is an outdated model trained with batch model while the teacher model in KD-self is the previous incremental model. We will compare their performance empirically in \nameref{sec:experiment} section. Given features of input data, the incremental model $M_t$ and the teacher model $Teacher_t$ make the predictions, as in line 4 and line 13. Then the incremental model $M_t$ is optimized by minimizing the loss function in Equation~\ref{loss_final}, as in line 14. The train process terminates when the model is trained with at least one epoch and KD loss stops decreasing, as in line 17.


\subsection{Data Module}
From data perspective, one straightforward way to tackle the catastrophic forgetting problems is to train the incremental model not only based on the new data, but also based on some selected historical data. We plan to implement a data reservoir to provide proper training data for incremental training. Some proportion of data in the existing reservoir and the new data are interleaved to be the new reservoir. In this module, some problems are needed to look into, such as what and which proportion of data in the existing reservoir should be kept. The implementation of data module is not finished for now and will be a part of future work to accomplish our framework.


\begin{algorithm}[H]
\caption{IncCTR}
\label{algo_incCTR}
\begin{algorithmic}[1]
    \Require features of incoming data $X_t$; labels of incoming data $Y_t$; existing model $M_{t-1}$; Teacher Model $Teacher_t$; existing policy of feature assignment $F_{t-1}$; frequency of existing features $S_{t-1}$
    \Ensure Model $M_t$
    \State \textbf{Initialize}: $L=MaxInt$; $ep$=0; EPOCH=1
    \State $\Theta_{t} \leftarrow $ \{Algorithm~\ref{algo_feature} $\cup$ network parameters\}
    \If{Train with KD}
    \State $Y_{soft} \leftarrow Inference(Teacher_t, X_t)$
    \State $M_t = Train\_with\_KD(X_t, Y_t, Y_{soft},\Theta_{t})$
    \Else
    \State $M_t = Train\_with\_FT(X_t, Y_t, \Theta_{t})$
    \EndIf
    \State \textbf{Return}: $M_t$
    \State 
    \State \textbf{\textit{Train\_with\_KD}}:
    \While{\textit{stopping\ criteria} not satisfied}
    \State $\hat{Y}_t = Inference(M_t, X_t)$
    \State $M_t \leftarrow \mathop{\arg\min}\limits_{\Theta_{t}}\left(\lambda\mathcal{L}_{CE}(Y_t, \hat{Y}_t) + \mathcal{L}_{KD}(\hat{Y}_t, Y_{soft}) + \mathcal{R}\right)$
    \State ep += 1
    \EndWhile
    \State \textit{stopping\ criteria}: $\mathcal{L}_{KD}(\hat{Y}_t, Y_{soft})$ increases and $ep$ >= EPOCH
    \State 
    \State \textbf{\textit{Train\_with\_FT}}:
    \While{\textit{stopping\ criteria} not satisfied}
    \State $\hat{Y}_t = Inference(M_t, X_t)$
    \State $M_t \leftarrow \mathop{\arg\min}\limits_{\Theta_{t}}\left(\mathcal{L}_{CE}(Y_t, \hat{Y}_t) + \mathcal{R}\right)$
    \State ep += 1
    \EndWhile
    \State \textit{stopping\ criteria}:\;$ep$ >= EPOCH
 \end{algorithmic}
\end{algorithm}

\section{Experiment}
\label{sec:experiment}
In this section, we conduct experiments on a public benchmark and a private dataset, aiming to answer the following research questions:
\begin{itemize}
    \item \textbf{RQ1}: What is the performance of IncCTR compared to training with batch mode?
    \item \textbf{RQ2}: What are the contribution of different modules in IncCTR framework? 
    \item \textbf{RQ3}: How efficient is IncCTR compared to training with batch mode?
\end{itemize}
\subsection{Dataset}
To evaluate the effectiveness and efficiency of the proposed IncCTR framework, we conduct extensive experiments both on a public benchmark and a private dataset. 
\begin{itemize}
\item Criteo\footnote{http://labs.criteo.com/downloads/download-terabyte-click-logs/}.
This dataset is used to benchmark algorithms for click-through rate (CTR) prediction. It consists of 24 days' consecutive traffic logs from Criteo, including 26 categorical features and 13 numerical features, and the first column as label indicating whether the ad has been clicked or not. 
\item HuaweiApp. In order to demonstrate the performance of the proposed method in real industrial tasks, we conduct offline experiments on a commercial dataset. HuaweiApp contains 60 consecutive days of click logs collected from Huawei AppGallery with user consent, consisting of app features, anonymous user features and context features.
\end{itemize}
For the ease of reproducing our experimental results, we present the details of data processing on Criteo data. In a nutshell, we follow~~Kaggle Champion\footnote{http://https//www.kaggle.com/c/criteo-display-ad-challenge/discussion/10555\label{fn:kaggle}} and~\cite{PNN}, which involving data sampling, discretization and feature filtering. We do not give details for processing HuaweiApp dataset for commercial reasons, but the procedure is similar.
\begin{itemize}
    \item Data sampling: Considering data imbalance (only 3\% samples are positive), similar to ~\cite{pin}, we apply negative down sampling to keep the positive ratio close to 50\%.
    \item Discretization: Both categorical and numerical features are existing in Criteo. However the distribution of two kinds of features is quite different intrinsically \cite{PNN}. In most of recommendation models, numerical features are transformed to categorical features through bucketing or logarithm. Following that, we use logarithm\textsuperscript{\ref{fn:kaggle}} as the discretization method in the formulation:
    \begin{equation}
        v\gets floor(log(v)^2)
    \end{equation} 
    \item Feature filtering: Infrequent features are usually not very informative and may be noisy, so that it is hard for models to learn such features well. Therefore, features in a certain field appearing less than 20 times are set a dummy feature \textit{Others}, following~\cite{PNN}. 
\end{itemize}

\begin{table*}[]
\caption{Overall performance comparison between IncCTR and training with batch mode over consecutive days on Criteo and HuaweiApp datasets. Mean AUC and Logloss over consecutive days are reported. The underlined numbers represent the performance of the best variant of IncCTR. Beside performance, average epochs and average train time (sec) for updating a new model are also presented for efficiency comparison.}
\label{tab:overall-performance}
\begin{tabular}{|l|lllll|lllll|}
\hline
\multicolumn{1}{|c|}{}                        & \multicolumn{5}{c|}{Criteo}                                                                                                              & \multicolumn{5}{c|}{HuaweiApp}                                                                                                            \\ \cline{2-11} 
\multicolumn{1}{|c|}{\multirow{-2}{*}{model}} & AUC                                          & Logloss                                      & avg epochs & avg time (s) & Impr. & AUC                                          & Logloss                                      & avg epochs & avg time (s) & Impr. \\ \hline
Batch-0                                       & 0.7977                                       & 0.5438                                       & 18         & 32694.66          & 0.06\%    & 0.8543                                       & 0.0859                                       & 17.34      & 91711.26          & -0.14\%   \\
Batch-1                                       & 0.7956                                       & 0.5464                                       & 9          & 16347.33          & 0.33\%    & 0.8532                                       & 0.0861                                       & 8.67       & 45855.63          & -0.01\%   \\
Batch-2                                       & 0.7946                                       & 0.5476                                       & 9          & 15907.42                 & 0.45\%    & 0.8523                                       & 0.0863                                       & 7.7          & 39697.77                 & 0.09\%    \\
Batch-3                                       & 0.7939                                       & 0.5485                                       & 8          & 14020.88                & 0.54\%    & 0.8514                                       & 0.0866                                       & 7.8          & 44587.14                 & 0.20\%    \\
Batch-4                                       & 0.7932                                       & 0.5492                                       & 10          & 17718.42                 & 0.63\%    & 0.8496                                       & 0.0870                                       & 7.57          & 43478.67                 & 0.41\%    \\
Batch-5                                       & {\color[HTML]{000000} 0.7925}                & {\color[HTML]{000000} 0.5501}                & 9          & 15853.46                 & 0.72\%    & {\color[HTML]{000000} 0.8475}                & {\color[HTML]{000000} 0.0874}                & 7.73          & 39030.66                 & 0.66\%    \\ \hline
IncCTR-Fine-tune                                     & {\color[HTML]{000000} {\ul \textbf{0.7982}}} & {\color[HTML]{000000} {\ul \textbf{0.5431}}} & {\ul 1}    & {\ul 258.48}      & {\ul 0}   & {\color[HTML]{000000} 0.8527}                & {\color[HTML]{000000} 0.0861}                & 1          & 167.35            & 0.05\%    \\
IncCTR-KD-batch                                      & {\color[HTML]{000000} 0.7982}                & {\color[HTML]{000000} 0.5442}                & 1.18       & 265.32            & 0         & {\color[HTML]{000000} {\ul \textbf{0.8531}}} & {\color[HTML]{000000} {\ul \textbf{0.0861}}} & {\ul 1.25} & {\ul 213.75}      & {\ul 0}   \\
IncCTR-KD-self                                       & {\color[HTML]{000000} 0.7981}                & {\color[HTML]{000000} 0.5462}                & 1          & 261.66            & 0.01\%    & {\color[HTML]{000000} 0.8530}                & {\color[HTML]{000000} 0.0862}                & 1.20       & 198.66            & 0.01\%    \\ \hline
\end{tabular}
\end{table*}

\subsection{Evaluation Protocols}

\textbf{Evaluation Metrics.} We adopt AUC (Area Under Curve) and logloss (cross-entropy) as our evaluation metrics which are widely used in CTR prediction models. Besides, it has been endorsed that an improvement of AUC and logloss at 0.1\% can be considered as significant for a CTR prediction model~\cite{cheng2016wide,deepfm,DCN}. \\
\textbf{Baseline.} Training model with batch mode is used as the baseline, to validate the effectiveness and efficiency of IncCTR. To further evaluate the impact on delay updating of models, we consider the baseline with different delay days. More specifically, Batch\_$i$ $(i=0,1,2,3,4,5)$ represents the baseline model with $i$ days' delay. \\
\textbf{Implementation Details.} As the focus of this work is to compare the effectiveness and efficiency of deep CTR models when training with batch mode and incremental training by IncCTR, we choose a popular deep CTR model DCN~\cite{DCN} for such comparison. Observations on other deep CTR models are similar. \\
To mimic the training procedure in industrial scenarios, experiments are conducted over consecutive days.
When training with batch mode, all the data in the fixed size window (i.e., data in size-$w$ window $[s,s+w)$, where $s\in[0,T-w]$) are utilized.
While training with incremental mode, only the data with the coming day (i.e., data in size-1 window $[s,s+1)$, where $s\in[w,T-1]$) is available.  
Warm-start is performed for the incremental models with a model trained with batch mode before the first incremental step. That is to say, we firstly train a warm-started model with batch mode on data in $[0,w)$, then train the first incremental model on the data of $w$-th day. We set $w=7, T=23$ for Criteo dataset and $w=30, T=59$ for HuaweiApp dataset. 




\subsection{Overall Performance (RQ1)}

Table \ref{tab:overall-performance} presents the overall performance comparison over consecutive test days. Models trained with batch mode with different delayed days Batch-$i$ ($i\in[1,5]$) are also compared here. Specifically, Batch-0 represents the model that is tuned on latest data (serving as validation) and then fine-tuned with latest data (serving as train data), namely training process is carried out twice. Batch-0 reaches the performance upper bound of batch mode, however, it is not feasible in practice as it doubles the training time of batch mode. Relative improvement of the best incremental model over the other models in terms of AUC is reported in ``Impr.'' column. From this table, we have the following observations.
\begin{itemize}   
    \item On Criteo dataset, incremental models achieve better effectiveness than baselines trained with batch mode which accelerates training procedure significantly. Specifically, incremental models outperform all the baselines with significant improvement (more than 0.1\%) over Batch-1 to Batch-5. Surprisingly, IncCTR achieves comparable performance to Batch-0 (which is an ideal model with upper bound performance).
    \item On HuaweiApp dataset, consistent results are acquired. Incremental models obtain comparable performance with baselines with huge efficiency improvement. Negligible decrease exists when comparing the performance of IncCTR with Batch-1 which can be ignored in practice. Unsurprisingly, Batch-0 utilizing the entire dataset outperforms the other models. Nonetheless, as stated earlier, Batch-0 is infeasible in practice as it doubles the training time in batch mode.
    \item On both datasets, severe performance degradation is observed as the delayed period extends, which calls for efficient training method. In industrial scenarios, updating delay of 1-3 days is a common phenomenon if the model is trained with batch mode in the single device which imputes to the enormous data volume, tedious preprocessing, cumbersome model structure and so on. Longer updating delay is also possible in some more complicated settings when multi-task or ensemble are needed. We can see that, with incremental learning method IncCTR, the  improvement of performance is quite significant when model updating delay occurs, for instance, AUC improves 0.6\% and 0.4\% when 5-day delay exists in HuaweiApp dataset and Criteo dataset (as shown in Figure~\ref{fig:mean_auc_huawei}).
\end{itemize}
\subsection{Ablation Studies (RQ2)}
\begin{figure}[h]
  \centering
  \includegraphics[width=0.8\linewidth]{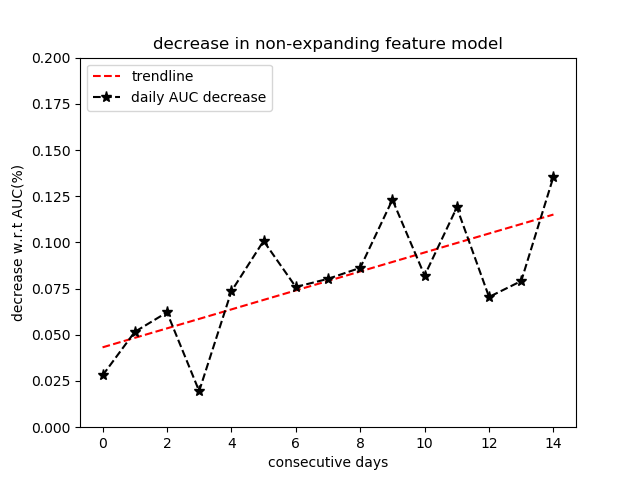}
  \caption{Performance decrease when new features are not expanded in Feature Module of IncCTR, over consecutive days on Criteo dataset.}
  \label{fig:decrease-non-expand-feat}
\end{figure}

To validate the contribution of feature module and model module in IncCTR, we do some ablation studies over such two modules.
\begin{itemize}
    \item Feature Module. In feature module, if the number of occurrences of a new feature from incoming data is above the threshold, we would assign an individual feature id to this feature. To verify the usefulness of the new feature expanding strategy, we conduct experiments to compare the effectiveness of whether expanding the new features during incremental learning over the two datasets. As shown in Table~\ref{tab:tab-fine-tune},  the consistent performance degradation demonstrates the necessity of expanding new features. 
    Specifically, because there are more new features per day on Criteo dataset, the strategy of new features expanding has greater impact on this dataset. Therefore, 
    the performance of IncCTR on Criteo dataset drops greater than that on HuaweiApp dataset when new features are not considered. 
    Besides, such decrease becomes more severe as the model keeps incrementally training, as shown in Figure~\ref{fig:decrease-non-expand-feat}, where more than $0.1\%$ performance decrease is observed eventually.
    \item Model Module. Fine-tune and knowledge distillation are applied in the model module. Compared with fine-tune, two KD methods (KD-batch and KD-self) achieve similar performance on Criteo and slight better performance on HuaweiApp dataset. This may because of too few new features emerging in new data. On average, only about 6\% new features arising each day in Criteo dataset, while similar phenomenon happens in HuaweiApp dataset. Comparing two KD methods, their performance are very close to each other, which suggests that KD-self is a better choice in practice as no extra computational resource is needed for training the teacher model.
\end{itemize}
\begin{table}[]
\caption{Feature Module: w/o new features V.S. with new features (AUC improvement).}
\label{tab:tab-fine-tune}
\begin{tabular}{lll}
\hline
model  & Criteo  & HuaweiApp \\ \hline
w/o new features & -       & -        \\ \hline
with new features   & +0.08\% & +0.055\% \\ \hline
\end{tabular}
\end{table}
\subsection{Efficiency (RQ3)}
Average epochs and training time (sec) of different models are summarized in Table~\ref{tab:overall-performance}. When training with batch mode, one epoch means going through over all data within the fixed-size window, while training with incremental mode, it means going through the incoming data. Tremendous advantage on efficiency is revealed, as we are able to get 60x and 270x improvement over average training time on Criteo and HuaweiApp dataset respectively, which is extremely helpful in practice.

\section{Conclusion}
\label{sec:conclusion}
In this paper, we propose the IncCTR, which is a practical incremental method to train deep CTR models. IncCTR includes data module, feature module and model module. Specifically, we propose new features expanding and initialization strategies in feature module, and propose several training algorithms in model module. Comprehensive experiments are conducted to demonstrate the effectiveness of the two modules in incCTR. Compared with conventional batch mode training, our method can achieve comparable performance in terms of AUC but with much better training efficiency, which is extremely helpful in practice.

There are two interesting directions for future study. One is investigating the novel approach to utilize history data to guarantee the stability of incremental training, such as reservoir. The other one is how to update (add or remove) the features incrementally and efficiently in the production environment of recommender system.

\newpage
\balance
\bibliographystyle{ACM-Reference-Format}
\bibliography{main}


\begin{thebibliography}{19}


\ifx \showCODEN    \undefined \def \showCODEN     #1{\unskip}     \fi
\ifx \showDOI      \undefined \def \showDOI       #1{#1}\fi
\ifx \showISBNx    \undefined \def \showISBNx     #1{\unskip}     \fi
\ifx \showISBNxiii \undefined \def \showISBNxiii  #1{\unskip}     \fi
\ifx \showISSN     \undefined \def \showISSN      #1{\unskip}     \fi
\ifx \showLCCN     \undefined \def \showLCCN      #1{\unskip}     \fi
\ifx \shownote     \undefined \def \shownote      #1{#1}          \fi
\ifx \showarticletitle \undefined \def \showarticletitle #1{#1}   \fi
\ifx \showURL      \undefined \def \showURL       {\relax}        \fi
\providecommand\bibfield[2]{#2}
\providecommand\bibinfo[2]{#2}
\providecommand\natexlab[1]{#1}
\providecommand\showeprint[2][]{arXiv:#2}

\bibitem[\protect\citeauthoryear{Cheng, Koc, Harmsen, Shaked, Chandra,
  et~al\mbox{.}}{Cheng et~al\mbox{.}}{2016}]%
        {cheng2016wide}
\bibfield{author}{\bibinfo{person}{Heng{-}Tze Cheng}, \bibinfo{person}{Levent
  Koc}, \bibinfo{person}{Jeremiah Harmsen}, \bibinfo{person}{Tal Shaked},
  \bibinfo{person}{Tushar Chandra}, {et~al\mbox{.}}}
  \bibinfo{year}{2016}\natexlab{}.
\newblock \showarticletitle{Wide {\&} Deep Learning for Recommender Systems}.
  In \bibinfo{booktitle}{\emph{DLRS@RecSys}}. ACM, \bibinfo{pages}{7--10}.
\newblock


\bibitem[\protect\citeauthoryear{Covington, Adams, and Sargin}{Covington
  et~al\mbox{.}}{2016}]%
        {dnnyoutube}
\bibfield{author}{\bibinfo{person}{Paul Covington}, \bibinfo{person}{Jay
  Adams}, {and} \bibinfo{person}{Emre Sargin}.}
  \bibinfo{year}{2016}\natexlab{}.
\newblock \showarticletitle{Deep neural networks for youtube recommendations}.
  In \bibinfo{booktitle}{\emph{Proceedings of the 10th ACM conference on
  recommender systems}}. ACM, \bibinfo{pages}{191--198}.
\newblock


\bibitem[\protect\citeauthoryear{Furlanello, Lipton, Tschannen, Itti, and
  Anandkumar}{Furlanello et~al\mbox{.}}{2018}]%
        {BANs}
\bibfield{author}{\bibinfo{person}{Tommaso Furlanello},
  \bibinfo{person}{Zachary~C. Lipton}, \bibinfo{person}{Michael Tschannen},
  \bibinfo{person}{Laurent Itti}, {and} \bibinfo{person}{Anima Anandkumar}.}
  \bibinfo{year}{2018}\natexlab{}.
\newblock \bibinfo{title}{Born Again Neural Networks}.
\newblock
\newblock
\showeprint[arxiv]{stat.ML/1805.04770}


\bibitem[\protect\citeauthoryear{Guo, Tang, Ye, Li, and He}{Guo
  et~al\mbox{.}}{2017}]%
        {deepfm}
\bibfield{author}{\bibinfo{person}{Huifeng Guo}, \bibinfo{person}{Ruiming
  Tang}, \bibinfo{person}{Yunming Ye}, \bibinfo{person}{Zhenguo Li}, {and}
  \bibinfo{person}{Xiuqiang He}.} \bibinfo{year}{2017}\natexlab{}.
\newblock \showarticletitle{Deepfm: a factorization-machine based neural
  network for ctr prediction}. In \bibinfo{booktitle}{\emph{IJCAI}}.
  \bibinfo{pages}{1725--1731}.
\newblock


\bibitem[\protect\citeauthoryear{He, Pan, Jin, Xu, Liu, Xu, Shi, Atallah,
  Herbrich, and Bowers}{He et~al\mbox{.}}{2014}]%
        {He2014Practical}
\bibfield{author}{\bibinfo{person}{Xinran He}, \bibinfo{person}{Junfeng Pan},
  \bibinfo{person}{Ou Jin}, \bibinfo{person}{Tianbing Xu}, \bibinfo{person}{Bo
  Liu}, \bibinfo{person}{Tao Xu}, \bibinfo{person}{Yanxin Shi},
  \bibinfo{person}{Antoine Atallah}, \bibinfo{person}{Ralf Herbrich}, {and}
  \bibinfo{person}{Stuart Bowers}.} \bibinfo{year}{2014}\natexlab{}.
\newblock \showarticletitle{Practical Lessons from Predicting Clicks on Ads at
  Facebook}. In \bibinfo{booktitle}{\emph{Eighth International Workshop on Data
  Mining for Online Advertising}}. \bibinfo{pages}{1--9}.
\newblock


\bibitem[\protect\citeauthoryear{Hinton, Vinyals, and Dean}{Hinton
  et~al\mbox{.}}{2015}]%
        {KD}
\bibfield{author}{\bibinfo{person}{Geoffrey~E. Hinton}, \bibinfo{person}{Oriol
  Vinyals}, {and} \bibinfo{person}{Jeffrey Dean}.}
  \bibinfo{year}{2015}\natexlab{}.
\newblock \showarticletitle{Distilling the Knowledge in a Neural Network}.
\newblock \bibinfo{journal}{\emph{CoRR}}  \bibinfo{volume}{abs/1503.02531}
  (\bibinfo{year}{2015}).
\newblock
\urldef\tempurl%
\url{http://arxiv.org/abs/1503.02531}
\showURL{%
\tempurl}


\bibitem[\protect\citeauthoryear{Li and Hoiem}{Li and Hoiem}{2016}]%
        {LWF}
\bibfield{author}{\bibinfo{person}{Zhizhong Li} {and} \bibinfo{person}{Derek
  Hoiem}.} \bibinfo{year}{2016}\natexlab{}.
\newblock \showarticletitle{Learning Without Forgetting}. In
  \bibinfo{booktitle}{\emph{Computer Vision - {ECCV} 2016 - 14th European
  Conference, Amsterdam, The Netherlands, October 11-14, 2016, Proceedings,
  Part {IV}}} \emph{(\bibinfo{series}{Lecture Notes in Computer Science})},
  \bibfield{editor}{\bibinfo{person}{Bastian Leibe}, \bibinfo{person}{Jiri
  Matas}, \bibinfo{person}{Nicu Sebe}, {and} \bibinfo{person}{Max Welling}}
  (Eds.), Vol.~\bibinfo{volume}{9908}. \bibinfo{publisher}{Springer},
  \bibinfo{pages}{614--629}.
\newblock


\bibitem[\protect\citeauthoryear{Lian, Zhou, Zhang, Chen, Xie, and Sun}{Lian
  et~al\mbox{.}}{2018}]%
        {xdeepfm}
\bibfield{author}{\bibinfo{person}{Jianxun Lian}, \bibinfo{person}{Xiaohuan
  Zhou}, \bibinfo{person}{Fuzheng Zhang}, \bibinfo{person}{Zhongxia Chen},
  \bibinfo{person}{Xing Xie}, {and} \bibinfo{person}{Guangzhong Sun}.}
  \bibinfo{year}{2018}\natexlab{}.
\newblock \showarticletitle{xDeepFM: Combining Explicit and Implicit Feature
  Interactions for Recommender Systems}.
\newblock \bibinfo{journal}{\emph{arXiv preprint arXiv:1803.05170}}
  (\bibinfo{year}{2018}).
\newblock


\bibitem[\protect\citeauthoryear{Liu, Tang, Chen, Yu, Guo, and Zhang}{Liu
  et~al\mbox{.}}{2019}]%
        {fgcnn}
\bibfield{author}{\bibinfo{person}{Bin Liu}, \bibinfo{person}{Ruiming Tang},
  \bibinfo{person}{Yingzhi Chen}, \bibinfo{person}{Jinkai Yu},
  \bibinfo{person}{Huifeng Guo}, {and} \bibinfo{person}{Yuzhou Zhang}.}
  \bibinfo{year}{2019}\natexlab{}.
\newblock \showarticletitle{Feature Generation by Convolutional Neural Network
  for Click-Through Rate Prediction}. In \bibinfo{booktitle}{\emph{The World
  Wide Web Conference, San Francisco, CA, USA, May 13-17}}.
  \bibinfo{publisher}{ACM}, \bibinfo{pages}{1119--1129}.
\newblock


\bibitem[\protect\citeauthoryear{McMahan, Holt, Sculley, Young, Ebner, Grady,
  Nie, Phillips, Davydov, Golovin, Chikkerur, Liu, Wattenberg, Hrafnkelsson,
  Boulos, and Kubica}{McMahan et~al\mbox{.}}{2013}]%
        {ftrl}
\bibfield{author}{\bibinfo{person}{H.~Brendan McMahan}, \bibinfo{person}{Gary
  Holt}, \bibinfo{person}{David Sculley}, \bibinfo{person}{Michael Young},
  \bibinfo{person}{Dietmar Ebner}, \bibinfo{person}{Julian Grady},
  \bibinfo{person}{Lan Nie}, \bibinfo{person}{Todd Phillips},
  \bibinfo{person}{Eugene Davydov}, \bibinfo{person}{Daniel Golovin},
  \bibinfo{person}{Sharat Chikkerur}, \bibinfo{person}{Dan Liu},
  \bibinfo{person}{Martin Wattenberg}, \bibinfo{person}{Arnar~Mar
  Hrafnkelsson}, \bibinfo{person}{Tom Boulos}, {and} \bibinfo{person}{Jeremy
  Kubica}.} \bibinfo{year}{2013}\natexlab{}.
\newblock \showarticletitle{Ad click prediction: a view from the trenches}. In
  \bibinfo{booktitle}{\emph{{ACM} {SIGKDD}}}.
\newblock
\urldef\tempurl%
\url{https://doi.org/10.1145/2487575.2488200}
\showDOI{\tempurl}


\bibitem[\protect\citeauthoryear{Qu, Cai, Ren, Zhang, Yu, Wen, and Wang}{Qu
  et~al\mbox{.}}{2016}]%
        {PNN}
\bibfield{author}{\bibinfo{person}{Yanru Qu}, \bibinfo{person}{Han Cai},
  \bibinfo{person}{Kan Ren}, \bibinfo{person}{Weinan Zhang},
  \bibinfo{person}{Yong Yu}, \bibinfo{person}{Ying Wen}, {and}
  \bibinfo{person}{Jun Wang}.} \bibinfo{year}{2016}\natexlab{}.
\newblock \showarticletitle{Product-Based Neural Networks for User Response
  Prediction}. In \bibinfo{booktitle}{\emph{{IEEE} 16th International
  Conference on Data Mining, {ICDM} 2016, December 12-15, 2016, Barcelona,
  Spain}}. ACM, \bibinfo{pages}{1149--1154}.
\newblock


\bibitem[\protect\citeauthoryear{Qu, Fang, Zhang, Tang, Niu, Guo, Yu, and
  He}{Qu et~al\mbox{.}}{2019}]%
        {pin}
\bibfield{author}{\bibinfo{person}{Yanru Qu}, \bibinfo{person}{Bohui Fang},
  \bibinfo{person}{Weinan Zhang}, \bibinfo{person}{Ruiming Tang},
  \bibinfo{person}{Minzhe Niu}, \bibinfo{person}{Huifeng Guo},
  \bibinfo{person}{Yong Yu}, {and} \bibinfo{person}{Xiuqiang He}.}
  \bibinfo{year}{2019}\natexlab{}.
\newblock \showarticletitle{Product-based Neural Networks for User Response
  Prediction over Multi-field Categorical Data}.
\newblock \bibinfo{journal}{\emph{{ACM} Trans. Inf. Syst.}}
  \bibinfo{volume}{37}, \bibinfo{number}{1} (\bibinfo{year}{2019}),
  \bibinfo{pages}{5:1--5:35}.
\newblock


\bibitem[\protect\citeauthoryear{Rendle}{Rendle}{2010}]%
        {fm}
\bibfield{author}{\bibinfo{person}{Steffen Rendle}.}
  \bibinfo{year}{2010}\natexlab{}.
\newblock \showarticletitle{Factorization machines}. In
  \bibinfo{booktitle}{\emph{ICDM}}. IEEE, \bibinfo{pages}{995--1000}.
\newblock


\bibitem[\protect\citeauthoryear{Wang, Fu, Fu, and Wang}{Wang
  et~al\mbox{.}}{2017a}]%
        {crossnet}
\bibfield{author}{\bibinfo{person}{Ruoxi Wang}, \bibinfo{person}{Bin Fu},
  \bibinfo{person}{Gang Fu}, {and} \bibinfo{person}{Mingliang Wang}.}
  \bibinfo{year}{2017}\natexlab{a}.
\newblock \showarticletitle{Deep \& cross network for ad click predictions}. In
  \bibinfo{booktitle}{\emph{ADKDD}}. ACM, \bibinfo{pages}{12}.
\newblock


\bibitem[\protect\citeauthoryear{Wang, Fu, Fu, and Wang}{Wang
  et~al\mbox{.}}{2017b}]%
        {DCN}
\bibfield{author}{\bibinfo{person}{Ruoxi Wang}, \bibinfo{person}{Bin Fu},
  \bibinfo{person}{Gang Fu}, {and} \bibinfo{person}{Mingliang Wang}.}
  \bibinfo{year}{2017}\natexlab{b}.
\newblock \showarticletitle{Deep {\&} Cross Network for Ad Click Predictions}.
  In \bibinfo{booktitle}{\emph{Proceedings of the ADKDD'17, Halifax, NS,
  Canada, August 13 - 17, 2017}}. \bibinfo{publisher}{{ACM}},
  \bibinfo{pages}{12:1--12:7}.
\newblock


\bibitem[\protect\citeauthoryear{Zhang, Du, and Wang}{Zhang
  et~al\mbox{.}}{2016}]%
        {Zhang2016Deep}
\bibfield{author}{\bibinfo{person}{Weinan Zhang}, \bibinfo{person}{Tianming
  Du}, {and} \bibinfo{person}{Jun Wang}.} \bibinfo{year}{2016}\natexlab{}.
\newblock \showarticletitle{Deep learning over multi-field categorical data}.
  In \bibinfo{booktitle}{\emph{European conference on information retrieval}}.
  Springer, \bibinfo{pages}{45--57}.
\newblock


\bibitem[\protect\citeauthoryear{Zhao, Xie, Jia, Qian, Ding, Sun, and Li}{Zhao
  et~al\mbox{.}}{2020}]%
        {DHPS_baidu}
\bibfield{author}{\bibinfo{person}{Weijie Zhao}, \bibinfo{person}{Deping Xie},
  \bibinfo{person}{Ronglai Jia}, \bibinfo{person}{Yulei Qian},
  \bibinfo{person}{Ruiquan Ding}, \bibinfo{person}{Mingming Sun}, {and}
  \bibinfo{person}{Ping Li}.} \bibinfo{year}{2020}\natexlab{}.
\newblock \showarticletitle{Distributed Hierarchical {GPU} Parameter Server for
  Massive Scale Deep Learning Ads Systems}.
\newblock \bibinfo{journal}{\emph{CoRR}}  \bibinfo{volume}{abs/2003.05622}
  (\bibinfo{year}{2020}).
\newblock
\urldef\tempurl%
\url{https://arxiv.org/abs/2003.05622}
\showURL{%
\tempurl}


\bibitem[\protect\citeauthoryear{Zhou, Mou, Fan, Pi, Bian, Zhou, Zhu, and
  Gai}{Zhou et~al\mbox{.}}{2018a}]%
        {dien}
\bibfield{author}{\bibinfo{person}{Guorui Zhou}, \bibinfo{person}{Na Mou},
  \bibinfo{person}{Ying Fan}, \bibinfo{person}{Qi Pi}, \bibinfo{person}{Weijie
  Bian}, \bibinfo{person}{Chang Zhou}, \bibinfo{person}{Xiaoqiang Zhu}, {and}
  \bibinfo{person}{Kun Gai}.} \bibinfo{year}{2018}\natexlab{a}.
\newblock \bibinfo{title}{Deep Interest Evolution Network for Click-Through
  Rate Prediction}.
\newblock
\newblock
\showeprint[arxiv]{stat.ML/1809.03672}


\bibitem[\protect\citeauthoryear{Zhou, Zhu, Song, Fan, Zhu, Ma, Yan, Jin, Li,
  and Gai}{Zhou et~al\mbox{.}}{2018b}]%
        {din}
\bibfield{author}{\bibinfo{person}{Guorui Zhou}, \bibinfo{person}{Xiaoqiang
  Zhu}, \bibinfo{person}{Chenru Song}, \bibinfo{person}{Ying Fan},
  \bibinfo{person}{Han Zhu}, \bibinfo{person}{Xiao Ma},
  \bibinfo{person}{Yanghui Yan}, \bibinfo{person}{Junqi Jin},
  \bibinfo{person}{Han Li}, {and} \bibinfo{person}{Kun Gai}.}
  \bibinfo{year}{2018}\natexlab{b}.
\newblock \showarticletitle{Deep interest network for click-through rate
  prediction}. In \bibinfo{booktitle}{\emph{SIGKDD}}. ACM,
  \bibinfo{pages}{1059--1068}.
\newblock


\end{thebibliography}




\end{document}